\documentstyle[aps,preprint]{revtex}
\begin{document}
\draft
\preprint{SOGANG-HEP-229/97}
\title{Spacetime Duality of BTZ Black Hole} 
\author{
Jeongwon Ho\footnote{E-mail address : 
jwho@physics.sogang.ac.kr},
Won T. Kim\footnote{E-mail address : 
wtkim@ccs.sogang.ac.kr}, and 
Young-Jai Park\footnote{E-mail address : 
yjpark@ccs.sogang.ac.kr}
}
\address{
Department of Physics and Basic Science 
Research Institute,\\
Sogang University, C.P.O.Box 1142, Seoul 
100-611, Korea\\
}
\date{December 1997}
\maketitle
\begin{abstract}
We consider the duality of the quasilocal black hole thermodynamics,
explicitly the quasilocal black hole thermodynamic first law, in
BTZ black hole solution as a special one of the three-dimensional
low energy effective string theory.
\end{abstract}
\bigskip
 
\newpage
  
\section{Introduction}
BTZ (Ba$\tilde{n}$ados, Teitelboim, 
and Zanelli) black hole [1], which is a solution for three-dimensional
general relativity with a negative cosmological constant, is a special
solution of the low energy effective string theory [2]. And it has a 
translational symmetry in angular coordinate $\phi $ direction. It follows
that there exists a corresponding dual solution, i.e., three-dimensional
charged black string [3]. The BTZ black hole and the 
three-dimensional charged black string have quite different geometries 
: the former does not have a curvature singularity, while the later has a 
timelike singularity. In this respect, it is of quite interest to study 
how physical quantities depending on geometry of the black hole
hole behave under the dual transformation.
Horne {\it et al.} have shown that for asymptotically flat solutions of 
three-dimensional low energy string theory, the duality of conserved 
quantities defined on asymptotic region is given in such a way that 
a mass is unchanged, while an axion charge and an angular momentum are 
interchanged each other [4].  
It has been also shown that the Hawking temperature and the horizon area 
(viz. entropy) of BTZ black hole are dual invariant [5].

However, there is lack of consideration for asymptotic behavior under the 
dual transformation. The BTZ black hole is asymptotically anti-de Sitter, 
while the three-dimensional charged black string is asymptotically flat. 
Thus, asymptotic flatness is in general not an appropriate assumption 
for the study of the duality. 
Instead, one may consider a finite spatial boundary so that one can study
the duality of quasilocal quantities independent of the asymptotic behavior 
of a given spacetime [6,7].

In this paper, we shall consider the duality of the quasilocal black hole 
thermodynamic first law of the BTZ black hole. In section II, dual solutions
of the BTZ black hole and its background spacetime are discussed.
The duality of the quasilocal black hole thermodynamic first law is studied
in section III. 

\section{BTZ Black Hole and Its Dual Solution}
Consider a three-dimensional manifold $M$, which has a finite timelike spatial 
boundary $\Sigma^r$ as well as two spacelike boundaries (initial and final ones
denoted by $\Sigma_{t^{\prime}}$ and $\Sigma_{t^{\prime \prime}}$, 
respectively). The boundary of $\Sigma_t$, which is denoted by $S^r_t$, is 
given by the intersection space of $\Sigma_t$ and $\Sigma^r$.
We assume that $\Sigma_t$ is orthogonal to $\Sigma^r$. The orthogonality 
means that on the boundary $\Sigma^r$, the timelike unit normal $u^a$ to 
$\Sigma_t$ and the spacelike unit normal $n^a$ to  $\Sigma^r$ satisfy the 
relation $u^a n_a|_{ \Sigma^r} =0 $. The induced metrics defined on 
$\Sigma_t$, $\Sigma^r$ and $S^r_t$ are denoted by $h_{ab}$, $\gamma_{ab}$ 
and $\sigma_{ab}$, respectively.

The three-dimensional low energy effective string action [3] is given by
\begin{eqnarray}
I & = & \frac{1}{2\pi} \int_{M}d^3x \sqrt{-g}
      \Phi \left[R+ \Phi^{-2} (\nabla \Phi )^2
     -\frac{1}{12}H^2 + \frac{4}{l^2}\right]
\nonumber \\ 
    && - \frac{1}{\pi} \int_{\Sigma_{t^{\prime}}
}^{\Sigma_{t^{\prime\prime}}} 
d^2x \sqrt{h} \Phi K 
     - \frac{1}{\pi} \int_{\Sigma^r} d^2x \sqrt{
-\gamma} \Phi \Theta~,
\end{eqnarray}
where the value $-(1/2)\ln\Phi$ describes a dilaton field and $H$ is a 
three-form field strength of an antisymmetric two form field $B$. 
Comparing with the three-dimensional 
general relativity, $l^{-2}$ 
can be interpreted as the negative cosmological 
constant $ - \Lambda$ [2]. 
In the boundary terms of the eq.(1), $K$ and $\Theta $
are traces of extrinsic curvatures of 
$\Sigma_{t}$ and $\Sigma^r$ as embedded in the 
three-dimensional spacetime $M$,
$K_{ab}=-h_{a}^{c}\nabla_{c} u_{b}$ 
and $\Theta_{ab}= -\gamma_{a}^{c} \nabla_{c}  
n_{b}$, respectively.
The boundary terms are involved 
such that when one applies a solution of equations of
motion into the action and requires the boundary 
condition that the field variables be fixed on the 
boundaries, the action has an extremum value.

When the dilaton field is set by $\Phi = 1$ on equations of motion of the 
action (1), one can obtain the BTZ solution and the anti-symmetric field 
$B_{ab}$ as follow
\begin{eqnarray}
   ds^2 &=& - N^2(r) dt^2 + N^{-2}(r)dr^2 + r^2 \left(d{\phi} 
+ N^{\phi}(r)dt \right)^2~,
\nonumber \\
   N^2(r) &=& - M + \frac{r^2}{l^2} + \frac{J^2}{4r^2}~,~~
   N^{\phi}(r) = - \frac{J}{2r^2} + \frac{J}{2r^2_+}~,
\nonumber \\
 B_{\phi t} &=& \frac{r^2}{l} - \frac{r^2_+}{l}~,
\end{eqnarray}
where all other components of the antisymmetric 
field $B_{ab}$ are zero. 
In eq.(2), there exist two coordinate singularities corresponding
to the outer and inner horizons from $N^2(r) =0$,
\begin{eqnarray}
r_{\pm} = \sqrt{Ml} \left[ \frac{1}{2}\left( 1 \pm \sqrt{1 
- \left(\frac{J}{Ml}\right)^2} \right) \right]^{1/2}~,~~|J| \leq Ml~.
\end{eqnarray}
We regard the black hole horizon as $r_{H}=r_{+}$. 
In the metric (2), we require the regularity of the
antisymmetric field and vanishing shift vector 
at the horizon as $N^{\phi}(r_H) =B_{\phi t}(r_H)=0$. 
The requirement of vanishing shift vector
at horizon is actually the same with doing a
coordinate transformation $\phi \rightarrow
\phi - \Omega_H t$ on a metric including
non-vanishing shift vector at horizon, 
where $\Omega_H $ is the 
angular velocity of the horizon [5]. We shall show that
this coordinate transformation insures that the dual antisymmetric 
field is also regular at horizon.

Since the BTZ metric and the anti-symmetric field are independent 
of the coordinate $\phi$, the solution has
a translational symmetry in the direction $\phi$.  
Thus there exists a corresponding 
dual solution by means of the transformation [8]
\begin{eqnarray}
g^d_{\phi\phi}& = &g^{-1}_{\phi\phi},~~ g_{\phi\alpha}^d 
 = B_{\phi\alpha}g^{-1}_{\phi\phi},
\nonumber \\
g^d_{\alpha\beta}& = &g_{\alpha\beta}-(g_{\phi\alpha}
g_{\phi\beta}
- B_{\phi\alpha}B_{\phi\beta})
g^{-1}_{\phi\phi},
\nonumber \\
B_{\phi\alpha}^d &=& g_{\phi\alpha}g^{-1}_{\phi\phi},~~
B^d_{\alpha\beta} = B_{\alpha\beta}
- 2g_{\phi [\alpha} B_{\beta ] \phi}g^{-1}_{\phi\phi},
\nonumber \\
\Phi^{d} &=& g_{\phi \phi} \Phi  ,
\end{eqnarray}
where $\alpha$, $\beta$ run over all directions 
except $\phi$. The corresponding dual solution is obtained by
\begin{eqnarray}
   ds^2_d &=& -N^2(r)dt^2 + N^{-2}(r)dr^{2}+ \frac{1}{r^2}\left(d\phi 
+ N^{\phi}_d(r) dt \right)^{2}~,
\nonumber \\
   N^{\phi}_{d}(r) &=& B_{\phi t}(r) = \frac{r^2}{l} - \frac{r^2_+}{l}~,
\nonumber \\
   B_{\phi t}^{d}(r) &=& N^{\phi}(r) = -\frac{J}{2r^2} + \frac{J}{2r^2_+}~, ~~
   \Phi_d(r) = r^2~.
\end{eqnarray}
Note that the dual solution has the horizon at the same position $r=r_+$
and the dual shift vector $N^{\phi}_d$ and the dual anti-symmetric field
$B^d_{\phi t}$ are also regular on the horizon. We set the coordinate $\phi$ 
to a compact one such as $\phi$ is periodically identified 
$\phi \sim \phi + 2\pi$. Then dual solutions represent 
the same conformal field theories [4]. 

The three-dimensional 
black string solution can be obtained from the dual solution (5) 
by making the coordinate transformation $t=l(r_+^2 - r_-^2)^{-1/2}({\tilde t}
+ {\tilde \phi})$, $\phi = (r_+^2 - r_-^2)^{1/2}{\tilde \phi}$, $r^2 =
l{\tilde r}$, and by the identification of parameters ${\cal M}=r_+^2/l$,
${\cal Q} = -J/2$ [2,5] as follows
\begin{eqnarray}
d{\tilde s}^2 &=& - \left( 1- \frac{{\cal M}}{{\tilde r}} \right)d{\tilde t}^2
+ \left( 1- \frac{{\cal Q}^2}{{\cal M}{\tilde r}} \right)d{\tilde \phi}^2
+\left( 1- \frac{{\cal Q}^2}{{\cal M}{\tilde r}} \right)^{-1}
\left( 1- \frac{{\cal M}}{{\tilde r}} \right)^{-1}\frac{l^2d{\tilde r}^2}{
4{\tilde r}^2},
\nonumber \\
{\tilde B}_{{\tilde \phi}{\tilde r}}&=& \frac{{\cal Q}}{{\tilde r}}
- \frac{{\cal Q}}{{\cal M}}, ~~~ {\tilde \Phi} = l{\tilde r}.
\end{eqnarray}
In the above coordinate transformation, we have rescaled the time coordinate.
It follows that surface gravity of the black string, i.e., 
Hawking temperature, does not agree with that of the BTZ black hole.
This is originated from the fact that the BTZ black hole is not asymptotically
flat and there is an ambiguity in normalization of the timelike
Killing field for the black hole. For simplicity, we will study the duality of
the BTZ black hole (2) comparing with the dual metric (5) instead of the 
black string (6).

On the other hand, since the BTZ metric (2) is asymptotically anti-de 
Sitter and the anti-symmetric field diverges in infinite
spatial region, quasilocal values of the BTZ black hole
are not well defined in the limit $r \rightarrow \infty$. However,
the unexpected divergence can be eliminated by introducing an
action of reference background $I_0$ and defining physical action $I_P$ as
$I_p \equiv I - I_0$. We choose the metric and the fields of the reference 
background spacetime as vacuum solution $M \rightarrow 0$, $J \rightarrow 0$  
\begin{eqnarray}
ds^2_0 &=&  -\frac{r^2}{l^2}dt^2 + \frac{l^2}{r^2}dr^2 + r^2 d\phi^2~,
\nonumber \\
\Phi_0 &=&1~,~~ B_{\phi t}^0 =r^2/l~.
\end{eqnarray}
Performing the dual transformation (4), one can easily 
obtain its dual solution 
as follows
\begin{eqnarray}
ds^2_{0d} &=& -\frac{r^2}{l^2}dt^2 + \frac{l^2}{r^2}dr^2 + \frac{1}{ r^2} 
\left(d\phi + \frac{r^2}{l}dt \right)^2~,
\nonumber \\
\Phi_0 ^d &=& r^2~,~~ B_{\phi t}^{0d} = 0~.
\end{eqnarray}

\section{Duality of the Quasilocal Thermodynamic First Law}
For the case that a black hole is 
embedded in a finite cavity, the quasilocal
thermodynamic first law is given by an integration form [7] as follows
\begin{eqnarray}
\delta S_{BH} = \int_{S^r_t} d\phi ~ \beta  
\left[\delta 
{\cal E} - \omega^a \delta {\cal J}_a  
+ V_a \delta Q^a +  (s^{ab}/2) \delta 
\sigma_{ab} + {\cal Y}\delta\Phi
\right]^{cl}_0,
\end{eqnarray}
where $\beta=\int N d\tau $ is the inverse 
temperature defined on the finite spatial 
boundary $\Sigma^r$ and $N\omega^a = N^a$, $NV_a =B_{at}$.
The superscript `$cl$' and the subscript `$0$' denote the BTZ black hole 
solution (2) and the reference background solution (8), respectively. 
${\cal E} $, ${\cal J}_a $, $Q^a$, $s^{ab}$, and ${\cal Y}$
are quasilocal surface energy, momentum, axion charge, stress and dilaton 
pressure densities defined by
\begin{eqnarray}
 {\cal E} &=& - \frac{\sqrt{\sigma}}{\pi}
\left(n^a \nabla_a \Phi -\Phi k \right),~
 {\cal J}_a =  \frac{2\sqrt{\sigma}}{\sqrt{h}} 
n_c \sigma_{ad} P^{cd},~ 
Q^a =  \frac{2\sqrt{\sigma}}{
\sqrt{h}}P_B^{ab}n_b ,
\nonumber \\
s^{ab} &=& \frac{\sqrt{\sigma}}{\pi}\Bigl[
\sigma^{ab}n^c \nabla_c \Phi + \Phi [k^{ab}
-\sigma^{ab}(k - n^c a_c)] \Bigr],
\nonumber \\
\cal{Y} &=& \frac{\sqrt{\sigma}}{\pi} \Bigl[
\Phi^{-1} n^c \nabla_c \Phi -(k - n^c a_c) \Bigr],
\end{eqnarray}
respectively, where $k$ is the trace of the extrinsic curvature
as embedded in $\Sigma_t$, $k_{ab} =
-\sigma^c_a D_c n_b$, and $D_c$ is the covariant
derivative on $\Sigma $. $a^c = u^a \nabla_a u^c = N^{-1} h^{ac}
\nabla_a N $ is an acceleration of the timelike unit 
normal $u^c$. The quantities in eq.(10) are 
called extensive variables which are composed 
by intensive variables, e.g., the lapse function 
and the shift vector. 

We are now ready to study 
the duality of the quasilocal black hole thermodynamic
first law (9).
Firstly, from eqs.(2) and (5), we can easily check that Hawking temperature 
$T_H $ is dual invariant
\begin{eqnarray}
T_H = T_H^d 
= \left. \frac{1}{4\pi}(N^2)^{\prime} 
\right|_{r=r_H} = \frac{r_+^2 - r_-^2}{2\pi r_+ l^2},
\end{eqnarray}
where $\prime$ denotes differentiation  with respect to the radial coordinate
$r$.  Moreover, since the lapse function $N$ is dual invariant, Tolman 
temperature $T_H/N(r) = T(r)$ [9], which is red-shifted 
temperature from the horizon to the finite spatial boundary, is also dual 
invariant, $T(r) = T(r)^d$.

For the string action (1) containing non-minimally coupled
scalar field, the black hole entropy involves effect of the scalar field
as follows [7,10]
\begin{eqnarray}
S_{BH} \approx  - \frac{1}{\pi}\int_{\Sigma^{r_H}} 
d\tau d\phi N\sqrt{\sigma}
[\Phi \Theta - n^a \partial_a \Phi]_{cl},
\end{eqnarray}
The expression for the black hole entropy (12) is a generalization
of that of the Einstein gravity, which is obtained from the eq.(12) as we set 
$\Phi = 1$. It can be seen that
the black hole entropy (12) satisfies the perimeter 
law, which is the (2+1)-dimensional version of the area law [1], 
and is dual invariant
\begin{eqnarray}
S_{BH}= S_{BH}^d
= 2 \cdot 2\pi r_H.
\end{eqnarray}

Now, let us consider
the duality of quasilocal densities. The physical quasilocal 
surface energy density ${\cal E}_p$,
the momentum density ${\cal J}_{\phi p}$, 
and the axion charge density $Q^{\phi}_p$ 
and dual ones are given by
\begin{eqnarray}
{\cal E}_p &=& {\cal E}^d_p 
= - \frac{1}{\pi}\left[\sqrt{-M + \frac{r^2}{l^2}+ \frac{J^2}{4r^2}} 
- \frac{r}{l} \right],
\nonumber \\
{\cal J}_{\phi p} &=& - Q_{d p}^{\phi} 
= - \frac{J}{2\pi},~~
Q^{\phi}_p = - {\cal J}_{\phi p}^d 
= \frac{1}{\pi l}.
\end{eqnarray}
Thus, the duality of the quasilocal densities defined on the 
finite spatial boundary are appeared as 
the physical quasilocal surface energy density is 
unchanged, while the physical momentum and 
the axion charge densities are interchanged each other.
This is the same behavior with the result of the Ref.[4] in which
the authors have studied the duality of conserved charges defined at
asymptotic flat region.

Since the lapse function is unchanged, while the 
shift vector and the antisymmetric field are 
interchanged each other under the dual transformation, the quantities
$\omega^{\phi} = N^{\phi}N^{-1}$ and $V_{\phi} =
B_{\phi t}N^{-1}$ are also interchanged under the
transformation. As a result, 
the first three terms in the thermodynamic first law (9) 
are invariant under the 
dual transformation.
\begin{eqnarray}
\delta {\cal E}_p - \omega^{\phi} 
\delta {\cal J}_{\phi p}  
+ V_{\phi} \delta Q^{\phi}_p
=\delta {\cal E}^d_p - \omega^{\phi}_{d} 
\delta {\cal J}_{\phi p}^d  
+ V_{\phi}^d \delta Q^{\phi}_{d p}.
\end{eqnarray}

Finally, we would like to study the dual behavior of the physical
quasilocal surface stress $s^{ab}_p$ 
and the dilaton pressure densities ${\cal Y}$. 
Existence of these quantities are 
originated from the fact that we have concerned with
the finite spatial boundary $\Sigma^r$. For the BTZ black hole,
i.e., the three-dimensional case,
the quasilocal surface stress density is just the `pressure' 
on the boundary conjugate to the perimeter $2\pi r$. Thus, 
the physical surface
pressure density times  the perimeter and the physical dilaton 
pressure density times the dilaton field can be interpreted as work terms.
It can be seen that the physical work terms are dual 
invariant such as not each other, but altogether as follows
\begin{eqnarray}
&&(s^{ab}_p/2)\delta \sigma_{ab} + {\cal Y}_p \delta \Phi
= {\cal P}_p\delta ({\rm perimeter})
+ {\cal P}_{\Phi p} \delta \Phi
\nonumber \\
&=&(s^{ab}_{dp}/2) \delta \sigma_{ab}^d + {\cal Y}^d_p \delta \Phi^d 
= {\cal P}^d_p \delta ({\rm perimeter})^d + {\cal P}_{\Phi p}^d \delta \Phi^d 
\nonumber \\
&=&  \frac{1}{4\pi^2} \left[\left(\frac{2r}{l^2} - \frac{J^2}{2r^3}\right)
\left[-M + \frac{r^2}{l^2}+ \frac{J^2}{4r^2} \right]^{-1/2}
- \frac{2}{l} \right] \delta (2\pi r)
\end{eqnarray}
Note that in eq.(16), the physical pressure term indeed 
vanishes at asymptotic region.

As a result, we have explicitly
shown that according to the eqs.(11), (13), (15) 
and (16), the quasilocal black hole
thermodynamic first law (9) is invariant
under the dual transformation (4).

\section*{Acknowledgments}
We were supported by Basic Science Research Institute Program,
Ministry of Education, Project No. BSRI-97-2414.

\end{document}